\documentclass[prb,twocolumn,showpacs,floats,eqsecnum,amsmath,amssymb]{revtex4}
\usepackage[dvips]{graphicx}

\begin{document}

\title{Magneto shot noise in noncollinear diffusive spin-valves}

\author{B. Abdollahipour and M. Zareyan}

\affiliation{Institute for Advanced Studies in Basic Sciences,
45195-1159 Zanjan, Iran}

\date{\today}

\begin{abstract}

We develop a semiclassical Boltzmann-Langevin theory of the spin
polarized shot noise in a diffusive normal metal spin-valve
connected by tunnel contacts to ferromagnetic reservoirs with
noncollinear magnetizations. We obtain basic equations for
correlations of the fluctuating spin-charge distribution and
current density matrices by taking into account the spin-flip
processes and precession of the spin accumulation vector in the
normal metal. Applying the developed theory to a two terminal FNF
structure, we find that for a small spin-flip strength and a
substantial polarization of the terminals the shot noise has a
nonmonotonic variation with the angle between magnetization
vectors. While the shot noise is almost unchanged from the normal
structure value for parallel configuration and increases well
above the normal value for antiparallel configuration, it
suppresses substantially at an intermediate angle depending on the
ratio of the conductances of the N metal and the tunnel contacts.
We also demonstrated pronounced effects of the polarization and
the spin-flip scattering on the shot noise which reveals the
interplay between relaxation and precession of the spin
accumulation vector in the N metal.

\end{abstract}


\maketitle

\section{Introduction}

Spin-polarized transport in magnetoelectronic structures has
recently attracted an intense interest largely due to their
important technological applications such as non-volatile magnetic
random access memories (MRAMs), read heads for mass data storage
and high sensitive magnetic sensors
\cite{Meservey94,Daughton99,Parkin02}. The main effect is the
(giant) magnetoresistance \cite{Levy94,Gijs97} in magnetic
multilayers and spin-valves, {\it i.e.} the remarkable decrease in
the resistance when the orientation of the magnetization vectors
of the ferromagnetic regions change from antiparallel to parallel.
A spin-valve consists of two ferromagnetic leads as the spin
injector and detector, connected through a normal metal spacer
which serves for the spin accumulation. Recent interest in
transport with noncollinear (neither parallel nor antiparallel)
magnetizations was stimulated by the spin-transfer magnetization
torque \cite{Slonczewski95,Berger96}, which is essential for the
new devices such as the spin-flip \cite{Brataas01,Xia02} and spin
torque transistors \cite{Bauer03}.
\par
The most appropriate theoretical formalism for noncollinear
ferromagnet-normal metal structures is the semiclassical circuit
theory \cite{Brataas01} which is based on dividing of the
structure into the reservoirs, nodes and the junctions and
expressing the currents in terms of the scattering matrices of the
junctions and the isotropic electronic distributions inside the
nodes and reservoirs \cite{Nazarov94}. For the noncollinear
spin-transport the spin-charge current and the corresponding
distributions inside the nodes and reservoirs, as well as the
scattering matrices acquire the $2\times 2$ matrix characteristics
of the Pauli spin space. The semiclassical Boltzmann equations
provide the standard method to calculate the noncollinear
distribution matrix in a continuous bulk medium in the diffusive
limit. Magnetoelectronic circuit theory complemented by the
Boltzmann diffusion equations have been widely used to study
different magnetoelectronic DC effects \cite{Huertas00}. In spite
of this there has been little attention toward the fluctuations of
the noncollinear spin-polarized current. Time-dependent current
fluctuations at low temperatures, the so called shot noise,
provides valuable information about the transport process which
are not extractable from the average current
\cite{Blanter00,Nazarov03}. In magnetoelectronic structures, in
which the spin of electrons plays an essential rule, the shot
noise is expected to contain spin-resolved information, including
spin-dependent correlations and spin accumulation and relaxation.
This together with the importance of the noise in
magnetoelectronic devices in view of applications motivate
studying of the spin-polarized current fluctuations.
\par
The theory for current fluctuations in disorder conducting systems
can be formulated by including the Langevin sources of
fluctuations due to the random scattering from disorders and the
fluctuations of the distribution in the semiclassical Boltzmann
equations\cite{Nagaev92}. The resulting Boltzmann-Langevin (BL)
formalism has been used to calculate the shot noise in many
mesoscopic structures and its results are widely believed to be
identical with an ensemble average of the exact quantum results
\cite{Blanter00}. One important example is the universal one-third
suppression \cite{Steinbach96,Henny99} of the shot noise in a
diffusive normal metal, which was correctly predicted within the
BL approach \cite{Beenakker92,Nagaev92}. Recently spin-polarized
shot noise has been studied theoretically in normal metals
connected by the ferromagnetic terminals with collinear
magnetizations \cite{Mishchenko03,Belzig04,Zareyan05,Zareyan005}.
In Ref. \onlinecite{Zareyan05} a semiclassical BL theory of the
collinear spin-polarized current fluctuations in a diffusive
normal metal was developed. It was found that in a multi-terminal
spin-valve structure the shot noise and the cross correlations
measured between currents of two different ferromagnetic terminals
can deviate substantially from the unpolarized values, depending
on the spin accumulation and the spin-flip scattering strength. On
the other hand only a few works \cite{Tserkovnyak01,Braun06} have
been devoted to shot noise in noncollinear systems. Very recently
Braun {\it et.al} \cite{Braun06} showed that the
frequency-dependent shot noise in a quantum dot spin-valve with
noncollinear magnetizations of ferromagnets in the presence of an
external magnetic field can be used to detect single-spin dynamics
in the quantum dot.
\par
Tserkovnyak and Brataas \cite{Tserkovnyak01} extended the circuit
theory to obtain the Landauer-B\"{u}ttiker (LB) formula for the
shot noise of a FN contact with noncollinear magnetization. They
analyzed the angular dependence of the shot noise in a N metal
node connected to two F reservoirs for different types of
junctions between the reservoirs and the node. However, the effect
of the spin-flip scattering inside the N metal as well as the
spatial variation of the noncollinear spin accumulation vector,
which are essential in diffusive spin-valves with wire geometry,
are disregarded so far. These effects require an extension of the
BL method to include the diffusion and the relaxation of the
noncollinear spin accumulation inside the diffusive N metal. To
our knowledge there is no BL theory for shot noise of a
noncollinear spin-polarized transport. The aim of the present work
is to develop such a theory.
\par
In this paper we develop a semiclassical BL equation for the
noncollinear spin-polarized current fluctuations in the presence
of the spin-flip scattering. We obtain the basic diffusion
equations for the fluctuating spin-charge distribution and current
matrices which allows to calculate the mean current and the
correlations of the corresponding fluctuations in a diffusive
normal metal connected by tunnel contacts to several F reservoirs.
The developed BL equations are supplemented by the the generalized
LB formula for the shot noise at the contact
points\cite{Tserkovnyak01}.
\par
To illustrated the main behavior of the noncollinear shot noise in
magnetoelectronic systems we apply the developed BL equations to
calculate the Fano factor, defined as the ratio of the noise power
to average current, in a two terminal FNF structure. The
noncollinear orientation of the magnetizations causes a precession
of the spin accumulation vector through the N metal, which in
presence of the spin-flip scattering is associated with a damping
as a function of the distance from the F reservoirs. For a small
spin-flip intensity and a finite polarization of the tunnel
contacts the precession of the spin accumulation results in a
nonmonotonic angular dependence of the Fano factor. For a parallel
configuration Fano factor is almost the same as the the normal
state value but for antiparallel configuration it increases well
above this value due to a large spin accumulation in the N metal.
We find a substantial decrease of the Fano factor at an
intermediate orientation determined from the conductances of the N
metal and the tunnel contacts. Introducing the spin-flip
scattering as well as decreasing the spin polarization diminish
the nonmonotonic behavior. We present a full analysis of the
magneto shot noise which demonstrates the effects of the spin-flip
induced relaxation and the precession imposed by noncollinear
magnetizations.
\par
The paper is organized as follows. Section II devotes to introduce
the BL equation for noncollinear transport and derive diffusion
equations. In section III we calculate all possible correlations
of intrinsic fluctuations and Langevin sources. Section IV devotes
to calculate fluctuations and average of the current in double
barrier FNF system. We show results for charge current shot noise
in section V and finally give some conclusions in section VI.


\section{Boltzmann-Langevin equations for noncollinear transport }

In this section we develop a semiclassical BL formalism for
fluctuations of the noncollinear spin-polarized current in a
diffusive normal metal connected through tunnel contacts to a
number of ferromagnetic reservoirs. When the magnetization vectors
of the reservoirs have a noncollinear orientation, the spin
accumulation vector in the normal metal will also have a
noncollinear direction with respect to the quantization axis
($z$). The semiclassical electronic distribution is then
determined by a $2\times 2$ matrix in the Pauli spin space of the
form
\begin{equation}
\hat{f}(\mathbf{k},\mathbf{r},t)=\left(
\begin{array}{cc}
\begin{array}{c}
f^{\uparrow\uparrow }(\mathbf{k},\mathbf{r},t)
\end{array}
&
\begin{array}{c}
f^{\uparrow \downarrow }(\mathbf{k},\mathbf{r},t)
\end{array}
\\
\begin{array}{c}
\\
f^{\downarrow \uparrow }(\mathbf{k},\mathbf{r},t)
\end{array}
&
\begin{array}{c}
\\
f^{\downarrow\downarrow }(\mathbf{k},\mathbf{r},t)
\end{array}
\end{array}
\right). \label{disfun}
\end{equation}
The fluctuating matrix distribution function
$\hat{f}(\mathbf{k},\mathbf{r},t)=\bar{\hat{f}}(\mathbf{k},\mathbf{r})+
\delta\hat{f}(\mathbf{k},\mathbf{r},t)$, depends on the momentum
$\mathbf{k}$, the coordinate $\mathbf{r}$ and time $t$. It is
convenient to expand $\hat{f}$ into the charge and the spin vector
distributions in terms of the $2 \times 2$ unit matrix and the
Pauli matrices $\{1,\hat{\mbox{\boldmath $\sigma$}}\equiv
(\sigma_x,\sigma_y,\sigma_z)\}$. The matrices for two
distributions whose spin vector distributions are pointed in
opposite directions take the form
\begin{eqnarray}
\hat{f}^{\pm}(\mathbf{k},\mathbf{r},t)=f_{\text{c}}(\mathbf{k},\mathbf{r},t)
\hat{1}\pm \hat{\mbox{\boldmath $\sigma$}}\cdot\vec{f}_{\text{s}}
(\mathbf{k},\mathbf{r},t).\label{fpm}
\end{eqnarray}
Here $f_{\text{c}}=(f^{\uparrow\uparrow }+f^{\downarrow\downarrow
})/2$ is the charge or the spin independent part of the
distribution matrix and $\vec{f}_{\text{s}}$ the spin distribution
vector whose $\text{z}$-component
$f_{\text{s}z}=(f^{\uparrow\uparrow }-f^{\downarrow\downarrow
})/2$ is spin polarization along the quantization axis and the
other two components $f_{\text{s}x}$, $f_{\text{s}y}$ describe the
spin-polarization oriented perpendicular to the quantization axis.
Note that the two distributions $\hat{f}^{\pm}$ are transformed to
each other by making the replacement
$\vec{f}_{\text{s}}\rightarrow-\vec{f}_{\text{s}}$.
\par
The BL equation for the noncollinear distribution matrix
(\ref{disfun}) is written as
\begin{eqnarray}
\frac{d}{dt}\hat{f}^{+}=\hat{I}^{\text{imp}}[\hat{f}^+]+\hat{I}^{\text
{sf}} [\hat{f}^+,\hat{f}^-]+\hat{\xi}^{\text
{imp}}+\hat{\xi}^{\text {sf}}, \label{ble}
\end{eqnarray}
where $\hat{I}^{{\text {imp(sf)}}}$ is the collision integral for
normal spin-independent impurity (spin-flip) scattering and
$\hat{\xi}^{{\text {imp(sf)}}}$ the corresponding Langevin source
of the current fluctuations. The matrix collision integrals are
expressed as the following
\begin{eqnarray}
\hat{I}^{\text{imp}}[\hat{f}^+]=\int
d\mathbf{k'}W^{\text{imp}}(\mathbf{k},\mathbf{k'})
[\hat{f}^+(\mathbf{k'})-\hat{f}^+(\mathbf{k})], \label{iimp}
\\*
\hat{I}^{\text{sf}}[\hat{f}^+,\hat{f}^-]=\int
d\mathbf{k'}W^{\text{sf}}(\mathbf{k},\mathbf{k'})
[\hat{f}^-(\mathbf{k'})-\hat{f}^+(\mathbf{k})], \label{isf}
\end{eqnarray}
where $W^{\text{imp}}(\mathbf{k},\mathbf{k'})$ is the rate of the
impurity scattering in which an electron scatters from the state
with momentum $\mathbf{k}$ into $\mathbf{k'}$ without changing its
spin state. The spin-flip scattering rate
$W^{\text{sf}}(\mathbf{k},\mathbf{k'})$ describes the transition
from the state with momentum $\mathbf{k}$ and spin state
$|\hat{s},\pm\rangle$ to $\mathbf{k'}$ and the flipped spin state
$|\hat{s},\mp\rangle$, where $|\hat{s},\mp\rangle$ are the up and
down spin eigen states in the quantization axis parallel to the
local spin polarization vector
$\vec{f}_{\text{s}}=|\vec{f}_{\text{s}}|\hat{s}$. In writing the
expressions (\ref{iimp}) and (\ref{isf}) we assumed that
$W^{\text{imp(sf)}}(\mathbf{k},\mathbf{k'})=W^{\text{imp(sf)}}(\mathbf{k'},\mathbf{k})$,
which follows from the detail balance principle. We also ignored
dependence of the scattering rates on the spin state of electron.
\par
The Langevin sources are expressed in terms of the fluctuating
part of the matrix distribution as follows
\begin{eqnarray}
\hat{\xi}^{\text{imp}}=\int
d\mathbf{k'}W^{\text{imp}}(\mathbf{k},\mathbf{k'})
[\delta\hat{f}^+(\mathbf{k'})-\delta\hat{f}^+(\mathbf{k})],
\label{lsimp}
\\*
\hat{\xi}^{\text{sf}}=\int
d\mathbf{k'}W^{\text{sf}}(\mathbf{k},\mathbf{k'})
[\delta\hat{f}^-(\mathbf{k'})-\delta\hat{f}^+(\mathbf{k})].
\label{lssf}
\end{eqnarray}
For a diffusive normal metal we apply the standard diffusive
approximation. Assuming that all quantities are sharply peaked
around Fermi level, the momentum vector $\mathbf{k}$ is expressed
in terms of the energy $\varepsilon$ and the direction of the
Fermi momentum $\mathbf{n}$. Then the following relations are hold
for the elastic scattering of electrons by the normal impurities
and spin-flip disorders:
\begin{eqnarray}
W^{{\text {imp}}({\text {sf}})}(\mathbf{k},\mathbf{k}^{\prime}
)=\frac{2}{N_0}\delta (\varepsilon-\varepsilon^{\prime}) w^{{\text
{imp}}({\text {sf}})} (\mathbf{n},\mathbf{n} ^{\prime}),
\label{wnnp}
\end{eqnarray}
where $N_0$ is the density of states in the Fermi level. In the
diffusive regime the electronic distribution is weakly anisotropic
in the momentum space, and the distribution matrix can be
expanded  up to the linear term in  $\mathbf{n}$:
\begin{eqnarray}
\hat{f}(\mathbf{n},\varepsilon,\mathbf{r},t)=
\hat{f}_{0}(\varepsilon,\mathbf{r},t)
+\mathbf{n}.\hat{\mathbf{f}}_{1}(\varepsilon,\mathbf{r},t).
\label{sda}
\end{eqnarray}
Here the anisotropic part of the matrix distribution is determined
by the vector $\mathbf{\hat{f}}_{1}$ whose components are $2\times
2$ matrices in the spin space. The matrix current density is
expressed as
\begin{eqnarray}
\hat{\mathbf{J}}(\mathbf{r},t)=\bar{\hat{\mathbf{J}}}(\mathbf{r})
+\delta \hat{\mathbf{J}}(\mathbf{r},t)=(eN_0\emph{v}_F/6)\int
d\varepsilon \hat{\mathbf{f}}_{1}, \label{cd}
\end{eqnarray}
where $v_{F}$ is the Fermi velocity. The isotropic part
$\hat{f}_{0}$ determines the fluctuating electrochemical
potential:
\begin{eqnarray}
\hat{\varphi}(\mathbf{r},t)=\bar{\hat{\varphi}}(\mathbf{r})+
\delta\hat{\varphi}(\mathbf{r},t)= (1/e)\int d\varepsilon
\hat{f}_{0}. \label{echp}
\end{eqnarray}
Replacing expansion (\ref{sda}) in Eq. (\ref{ble}) we obtain the
diffusion equations for the current density and the potential
matrices, which read
\begin{eqnarray}&&
{\nabla}^2\bar{\hat{\varphi}}(\mathbf{r})=\frac{1}{\ell^{2}_{\text
{sf}}}\left(\bar{\hat{\varphi}}-\frac{Tr(\bar{\hat{\varphi}})}{2}\hat{1}\right),
\label{lapfi}
\\*&&
\nabla \cdot \hat{\mathbf{J}}=-\frac{e^2 N _{0}} {2\tau _{{\text
{sf}}}}\left(\hat{\varphi}-\frac{Tr(\hat{\varphi})}{2}\hat{1}\right)
+\hat{\mbox{\boldmath $\sigma$}} \cdot \mbox{\boldmath
$\zeta$},\label{divj}
\\*&&
\hat{\mathbf{J}}=-\mathcal{\sigma} \nabla \hat{\varphi} +
\hat{\mbox{\boldmath $\eta$}}. \label{j}
\end{eqnarray}
In these equations $\sigma= e^2 (N_{0}/2) D$ is the conductivity
and $\ell_{\text {sf}}=\sqrt{D\tau_{\text {sf}}}$ is the spin-flip
scattering length, where the diffusion constant is given by
$D=v_{F}^2\tau/3$. The impurity and the spin-flip relaxation times
are defined by the following relations
\begin{eqnarray}&&
\frac{\mathbf{n}}{\tau _{{\text {imp}}}}= \int
d\mathbf{n}^{\prime} w^{{\text {imp}}}(\mathbf{n},
\mathbf{n}^{\prime}) (\mathbf{n}-\mathbf{n}^{\prime}),
\label{tauim}
\\*&&
\frac{\mathbf{n}}{\tau_{{\text {sf}}}}=2\int d\mathbf{n}^{\prime}
w^{\text {sf}}(\mathbf{n},\mathbf{n}^{\prime}) (\mathbf{n}-
\mathbf{n}^{\prime}), \label{tausf}
\\*&&
\frac{1}{\tau}=\frac{1}{\tau _{{\text
{imp}}}}+\frac{1}{\tau_{{\text {sf}}}}.
\end{eqnarray}
The matrix Langevin current fluctuations in Eq. (\ref{j}) has the
form
\begin{eqnarray}
\hat{\mbox{\boldmath $\eta$}}= ev_{F}(N_{0}/2)\tau \int
(\hat{\xi}^{{\text {imp}}} + \hat{\xi}^{\text {sf}})\mathbf{n}
d\mathbf{n}d\varepsilon. \label{jcs}
\end{eqnarray}
Also,
\begin{eqnarray}
\hat{\mbox{\boldmath $\sigma$}} \cdot \mbox{\boldmath
$\zeta$}(\mathbf{r},t) =(eN_{0}/2)\int d\varepsilon
d\mathbf{n}\left(\hat{\xi}^{\text {sf}}-\frac{Tr(\hat{\xi}^{\text
{sf}})}{2}\hat{1}\right), \hspace{0.5cm}\label{isp}
\end{eqnarray}
is the matrix Langevin divergence term due to the non-conserved
nature of the spin-flip scattering. These two fluctuating Langevin
terms have zero average values.
\par
Note that we disregarded the explicit dependence of $\hat{f}$ on
time, given by the term $\partial/\partial t$ of $d/dt$ in the
left hand side of the BL equation, since we are interested in the
zero frequency noise power.  To obtain these equations we focused
on the realistic limit when $\tau _{{\text {imp}}} \ll \tau
_{{\text {sf}}}$ and ignored the effect of spin-flip scattering on
the conductivity of the diffusive normal metal.


\section{correlations of current fluctuations}

\subsection{Correlations of the Langevin sources of current fluctuations}

To calculate correlations between components of the fluctuating
Langevin terms  $\hat{\mbox{\boldmath $\eta$}}$ and
$\mbox{\boldmath $\zeta$}$ we refer again to the BL equation
(\ref{ble}). But, now we choose the spin quantization axis to be
parallel to the local mean spin distribution vector
$\bar{\vec{f}}_{\text{s}}$. Denoting the projection matrices along
$\bar{s}$ and two perpendicular unit vectors $\bar{s}_{\bot i}
(i=1,2)$, by $\hat{\upsilon}_{\bar{s}}$ ($=\hat{\mbox{\boldmath
$\sigma$}}\cdot\bar{s} $) and $\hat{\upsilon}_{\bar{s}_{\bot i}}$
respectively, the matrix distribution function can be expanded as
\begin{eqnarray}
\hat{f}^{\pm}=\frac{1}{2}(f_{\bar{s}_{+}}+f_{\bar{s}_{-}})\hat{1}\pm
\frac{1}{2}(f_{\bar{s}_{+}}-f_{\bar{s}_{-}})\hat{\upsilon}_{\bar{s}}\pm
\nonumber
\\*
\frac{1}{2}\sum_i(\delta f_{\bar{s}_{\bot i,+}}-\delta
f_{\bar{s}_{\bot i,-}})\hat{\upsilon}_{\bar{s}_{\bot
i}},\label{avfpm}
\end{eqnarray}
where the fluctuating transverse components of the spin
polarization vector have zero mean values.
\par
We substitute the fluctuating parts of the matrices (\ref{avfpm})
into the Langevin source terms in Eq. (\ref{ble}) and obtain the
result
\begin{eqnarray}
\hat{\xi}^{\text{imp(sf)}}&=&
\frac{1}{2}(\xi^{\text{imp(sf)}}_{\bar{s},+}+
\xi^{\text{imp(sf)}}_{\bar{s},-})\hat{1}+\frac{1}{2}(\xi^{\text{imp(sf)}}
_{\bar{s},+}- \nonumber
\\*&&
\xi^{\text{imp(sf)}}
_{\bar{s},-})\hat{\upsilon}_{\bar{s}}+\frac{1}{2}\sum_i(\xi^{\text{imp(sf)}}_{\bar{s}_{\bot
i,+}}-\xi^{\text{imp(sf)}} _{\bar{s}_{\bot
i,-}})\hat{\upsilon}_{\bar{s}_{\bot i}},\nonumber
\\*&&
\end{eqnarray}
where
\begin{eqnarray}&&
\xi^{\text{imp(sf)}}_{\hat{s},\alpha}=\int d\mathbf{k'}
[\delta{J}^{\text{imp(sf)}}_{\hat{s},(-)\alpha\alpha}(\mathbf{k'},\mathbf{k})-
\delta{J}^{\text{imp(sf)}}_{\hat{s},\alpha (-)
\alpha}(\mathbf{k},\mathbf{k'})], \nonumber
\\*&&
\delta{J}^{\text{imp(sf)}}_{\hat{s},\alpha\alpha^{\prime}}(\mathbf{k},\mathbf{k'})=
J^{\text{imp(sf)}}_{\hat{s},\alpha\alpha^{\prime}}(\mathbf{k},\mathbf{k'})-
\bar{J}^{\text{imp(sf)}}_{\hat{s},\alpha\alpha^{\prime}}(\mathbf{k},\mathbf{k'}),
\nonumber
\end{eqnarray}
and we have introduced the current in individual impurity
(spin-flip) scattering as
$J^{\text{imp(sf)}}_{\hat{s},\alpha\alpha^{\prime}}
(\mathbf{k},\mathbf{k'})=W^{\text{imp(sf)}}
(\mathbf{k},\mathbf{k'})f_{\hat{s},\alpha}(\mathbf{k})
\left[1-f_{\hat{s},\alpha^{\prime}}(\mathbf{k'})\right]$ for
component of spin in direction of $\hat{\mathbf{s}}$.
\par
Now we apply the central assumption of the BL approach and assume
that all scattering events are independent elementary processes
and thus the correlations of the associated currents fluctuations
obey the Poissonian relation:
\begin{eqnarray}
\langle\delta J^{\text{imp(sf)}}_{\hat{s}_i,\alpha_1
\alpha_2}(\mathbf{k}_1,\mathbf{k}_2, \mathbf{r},t) \delta
J^{\text{imp(sf)}}_{\hat{s}_j,\alpha_3 \alpha_4 }(\mathbf{k}_3
,\mathbf{k}_4,\mathbf{r}^{\prime},t^{\prime})\rangle \nonumber
\\*
=\delta_{ij}\delta _{\alpha_1\alpha_3}
\delta_{\alpha_2\alpha_4}\delta(\mathbf{k}_1-\mathbf{k}_3)
\delta(\mathbf{k}_2-\mathbf{k}_4) \nonumber
\\*
\times\delta(\mathbf{r}-\mathbf{r}^{\prime})\delta(t-t^{\prime})
{\bar J^{\text{imp(sf)}}_{\hat{s}_i,\alpha_1 \alpha
_2}(\mathbf{k}_1,\mathbf{k}_2,\mathbf{r},t)}, \label{poisson1}
\end{eqnarray}
and
\begin{eqnarray}
\langle\delta J^{\text{imp}}_{\hat{s}_i,\alpha_1
\alpha_2}(\mathbf{k}_1,\mathbf{k}_2, \mathbf{r},t) \delta
J^{\text{sf}}_{\hat{s}_j,\alpha_3 \alpha_4 }(\mathbf{k}_3
,\mathbf{k}_4,\mathbf{r}^{\prime},t^{\prime})\rangle=0
\hspace{0.5cm} \label{poisson2}
\end{eqnarray}
Using these relations we can calculate the correlations between
matrix elements of the Langevin sources
($\hat{\xi}^{\text{imp(sf)}}$), from that we obtain all the
possible correlations between components of the vector matrices
$\hat{\mbox{\boldmath $\eta$}}$ and $\mbox{\boldmath $\zeta$}$.
The results read
\begin{eqnarray}&&
\langle\eta_{l}^{\alpha\beta}(\mathbf{r},t)
\eta_{m}^{\alpha^{\prime}\beta^{\prime}}(\mathbf{r}^{\prime},t^{\prime})\rangle
=\frac{1}{2}\delta_{lm} \delta(\mathbf{r}-\mathbf{r}^{\prime})
\nonumber \delta(t-t^{\prime})\sigma\nonumber
\\*&&
\times\left[\left(\delta_{\alpha\beta}(\hat{\mbox{\boldmath
$\sigma$}}\cdot\bar{s})_{\alpha^{\prime}\beta^{\prime}}+
\delta_{\alpha^{\prime}\beta^{\prime}}(\hat{\mbox{\boldmath$\sigma$}}
\cdot\bar{s})_{\alpha\beta}\right)\sum_{\nu}\nu \Pi_{\bar{s},\nu
\nu}({\bf r}) \nonumber \right.
\\*&&
\left.+((\hat{\mbox{\boldmath
$\sigma$}}\cdot\bar{s})_{\alpha\beta}(\hat{\mbox{\boldmath
$\sigma$}}\cdot\bar{s})_{\alpha^{\prime}\beta^{\prime}}+
\delta_{\alpha\beta}\delta_{\alpha^{\prime}\beta^{\prime}})
\sum_{\nu} \Pi_{\bar{s},\nu \nu}({\bf r})\right],\nonumber\\*
\label{jcjc}
\end{eqnarray}
\begin{eqnarray}
\langle \zeta^{\gamma}(\mathbf{r},t) \zeta^{\gamma'}({\bf
r}^{\prime},t^{\prime})\rangle =
\delta(\mathbf{r}-\mathbf{r}^{\prime})\delta(t-t^{\prime})\sigma\nonumber
\\*
\times\frac{1}{D\tau_{\text{sf}}}\bar{s}^{\gamma}\bar{s}^{\gamma'}
\sum_{\nu}\Pi_{\bar{s},\nu-\nu}(\mathbf{r}),\label{isis}
\end{eqnarray}
\begin{eqnarray}
\langle\eta_{l}^{\alpha\beta}(\mathbf{r},t) \zeta^{\gamma}({\bf
r}^{\prime},t^{\prime})\rangle=0, \label{jcis}
\end{eqnarray}
where
\begin{equation}
\Pi_{\bar{s},\nu \nu^{\prime}}(\mathbf{r})= \int d\varepsilon
{\bar f}_{\bar{s},\nu}(\varepsilon,\mathbf{r}) [1-{\bar
f}_{\bar{s},\nu^{\prime}}(\varepsilon,\mathbf{r})]. \label{pi}
\end{equation}
Here ${\bar f}_{\bar{s},\nu }(\varepsilon,\mathbf{r})$ is the
isotropic part of the up and down spin components of the mean
distribution function with respect to the quantization axis
parallel to $\bar{\mathbf{s}}$. So, if we know the mean matrix
distribution function we can easily calculate these correlations.
It is a remarkable result that, just the longitudinal fluctuations
of the distribution function has non-vanishing contribution to
correlations and hence in shot noise. But, the transverse
fluctuations of distribution function, actually make no
contribution in the shot noise.


\subsection{Boundary conditions and correlations of intrinsic fluctuations}

The diffusion equations (\ref{lapfi}-\ref{j}) and Eqs.
(\ref{jcjc}-\ref{jcis}) are the extension of the BL equations
obtained in Ref. \onlinecite{Zareyan05} for the collinear
spin-polarized shot noise to the noncollinear magnetizations.
These equations are a complete set of equations, which have to be
implemented by the appropriate boundary conditions at the contacts
of the normal metal to the ferromagnetic terminals. For a FN
junction with a noncollinear magnetization the expression for the
matrix current reads \cite{Brataas01}
\begin{eqnarray}
e\bar{\hat{I}}^{\text{C}}&=& G^{\uparrow\uparrow}
\hat{u}^{\uparrow} \left(
\bar{\hat{f}}_{\text{F}}-\bar{\hat{f}}_{\text{N}} \right)
\hat{u}^{\uparrow} + G^{\downarrow\downarrow} \hat{u}^{\downarrow}
\left(\bar{\hat{f}}_{\text{F}}-\bar{\hat{f}}_{\text{N}}\right)
\hat{u}^{\downarrow} \nonumber
\\*&&
- G^{\uparrow \downarrow} \hat{u}^{\uparrow}
\bar{\hat{f}}_{\text{N}} \hat{u}^{\downarrow} - (G^{\uparrow
\downarrow})^{\ast} \hat{u}^{\downarrow} \bar{\hat{f}}_{\text{N}}
\hat{u}^{\uparrow} \, , \label{avcurrent}
\end{eqnarray}
where $\bar{\hat{f}}_{\text{N}}$ ($\bar{\hat{f}}_{\text{F}}$) is
the average matrix distribution function in the normal
(ferromagnetic) side of the contact, $G^{\uparrow\uparrow
,\downarrow\downarrow} = \frac{e^2}{h} \left[ M-\sum_{nm}
|r_{nm}^{\uparrow,\downarrow}|^2 \right]$ are the diagonal
components of the conductance matrix describing currents of
electrons with spin parallel (up) and antiparallel (down) to the
magnetization vector of F, where M is the number of transverse
modes in the contact channel and $r^{\uparrow}_{nm}$
($t^{\uparrow}_{nm}$) and $r^{\downarrow}_{nm}$
($t^{\downarrow}_{nm}$) are the spin up and down reflection
(transmission) coefficients in the basis where the
spin-quantization axis is parallel to the magnetization of F. The
mixing conductance $G^{\uparrow \downarrow}=\frac{e^2}{h} \left[
M-\sum_{nm} r^{\uparrow}_{nm} (r^{\downarrow}_{nm})^{\ast} \right]
$, describes the transfer of the spin-current perpendicular to the
magnetization of F. Here $\hat{u}^{\uparrow
(\downarrow)}=\frac{1}{2}\left[\hat{1}+(-)\hat{\mbox{\boldmath
$\sigma$}} \cdot {\bf \hat{m}}\right]$ are projection matrices in
the spin space in which $\hat{m}$ is the unit vector in direction
of magnetization vector.
\par
To write an expression for the fluctuations of the matrix current
(\ref{avcurrent}) we follow Beenakker and B\"{u}ttiker
assumption\cite{Beenakker92} and consider that the time-dependent
fluctuations have two contributions. The first contribution
$\delta\hat{I}(t)$ is the intrinsic fluctuations of the matrix
current, which is caused by randomness of the electron scatterings
from the junction. This is the only relevant term for the shot
noise of a single junction connecting two reservoirs which are
held at equilibrium. The second contribution can exist when the
distribution functions in N metal and/or the adjacent reservoirs
are fluctuating. For a contact between the N metal with
fluctuating part of the matrix distribution
$\delta\hat{f}_{\text{N}}$ and a F reservoir held at equilibrium,
the expression for the matrix current fluctuations has the form
\begin{eqnarray}
e\Delta\hat{I}^{\text{C}}=\delta\hat{I}-G^{\uparrow\uparrow
}\hat{u}^{\uparrow}\delta\hat{f}_{\text{N}}(\mathbf{r})\hat{u}^{\uparrow}-G^{\downarrow\downarrow
}\hat{u}^{\downarrow }\delta\hat{f}_{\text{N}}(\mathbf{r})
\hat{u}^{\downarrow }\nonumber
\\*
-G^{\uparrow \downarrow }\hat{u}^{\uparrow }\delta\hat{f}_{\text{N}}(\mathbf{r})\hat{u}%
^{\downarrow }-( G^{\uparrow \downarrow }) ^{\ast }\hat{u}%
^{\downarrow}\delta\hat{f}_{\text{N}}(\mathbf{r})\hat{u}^{\uparrow
}. \label{dcc}
\end{eqnarray}
The correlations of the intrinsic fluctuations for an arbitrary
type of a FN contact with a noncollinear magnetization was
calculated by Tserkovnyak and Brataas \cite{Tserkovnyak01} using a
generalized LB approach. Here we use their result and write the
correlations between different elements of the intrinsic matrix
current fluctuations $C^{\alpha\beta\alpha'\beta'}=\int
dt\langle\delta\hat{I}^{\alpha\beta}(t)\delta\hat{I}^{\alpha'\beta'}(0)
\rangle$, as follow
\begin{widetext}
\begin{eqnarray}
C^{\alpha\beta\alpha'\beta'}&=&\int d\varepsilon\left\{
\sum_{s_1s_2s_3s_4}\check{S}^{s_1s_2s_3s_4}\left[
\left(\hat{u}^{s_1}\left(\bar{\hat{f}}_{\text{N}}-\bar{\hat{f}}_{\text{F}}\right)\hat{u}^{s_2}\right)_{\alpha^{\prime}\beta}
\left(\hat{u}^{s_3}\left(\bar{\hat{f}}_{\text{N}}-\bar{\hat{f}}_{\text{F}}\right)\hat{u}^{s_4}\right)_{\alpha\beta^{\prime}}\right]
\nonumber\right.
\\*&&
\hspace{1cm}\left.+\sum_{s_1s_2}\hat{G}^{s_1s_2}\left[\left(\hat{u}^{s_1}\bar{\hat{f}}_{\text{N}}\right)_{\alpha^{\prime}\beta}
\left((\hat{1}-\bar{\hat{f}}_{\text{N}})\hat{u}^{s_2}\right)_{\alpha\beta^{\prime}}+\left(\bar{\hat{f}}_{\text{N}}\hat{u}^{s_2}\right)
_{\alpha^{\prime}\beta}\left(\hat{u}^{s_1}(\hat{1}-\bar{\hat{f}}_{\text{N}})\right)_{\alpha\beta^{\prime}}
\nonumber\right.\right.
\\*&&
\hspace{2.8cm}\left.\left.-\left(\hat{u}^{s_1}\bar{\hat{f}}_{\text{N}}\hat{u}^{s_2}\right)_{\alpha^{\prime}\beta}\left(\hat{1}-
\bar{\hat{f}}_{\text{F}}\right)_{\alpha\beta^{\prime}}-\left(\bar{\hat{f}}_{\text{F}}\right)_{\alpha^{\prime}\beta}\left(\hat{u}^{s_1}(\hat{1}-
\bar{\hat{f}}_{\text{N}})\hat{u}^{s_2}\right)_{\alpha\beta^{\prime}}\right]\right\}
\label{intcorrel} \,.
\end{eqnarray}
\end{widetext}
In this equation $\hat{G}^{s_1s_2}=\left[
M-\sum_{nm}r_{nm}^{s_1}(r_{mn}^{s_2})^{\ast}\right]$ and
$\check{S}^{s_1s_2s_3s_4}=\frac{e^2}{h}
\left[M-\sum_{nmn^{\prime}m^{\prime}}r_{nm}^{s_1}(r_{mn^{\prime}}^{s_2})^{\ast}
r_{n^{\prime}m^{\prime}}^{s_3}(r_{m^{\prime}n}^{s_4})^{\ast}\right]$
are elements of the conductance matrix $\hat{G}$ and the shot
noise matrix $\check{S}$ respectively. The general expressions of
these matrices are considerably simplified for a tunnel barrier,
for which an expansion in terms of the small transmission
coefficients $\delta r_{nm}^{s}=1-r_{nm}^{s} $, leads to the
following results\cite{Tserkovnyak01}
\begin{eqnarray}&&
\hat{G}=\frac{G}{2}\left(
\begin{array}{cc}
1+P & 1\\
1 & 1-P
\end{array}
\right)\label{conductance} ,
\\*&&
\check{S}=\frac{G}{2}\left(
\begin{array}{cccc}
2+2P & 2+P & 2+P & 2 \\
2+P & 2 & 2 & 2-P\\
2+P & 2 & 2 & 2-P\\
2 & 2-P & 2-P & 2-2P
\end{array}
\right)\label{noise}.\hspace{0.3cm}
\end{eqnarray}
Where we have the total conductance
$G=G^{\uparrow\uparrow}+G^{\downarrow\downarrow}$, the
polarization
$P=(G^{\uparrow\uparrow}-G^{\downarrow\downarrow})/G$, $Re
G^{\uparrow\downarrow}=G/2$ and $Im G^{\uparrow\downarrow}=0$ for
the tunnel contact. Thus by knowing the average distribution
function matrices inside the connected N and F metals, we can
calculate all possible correlations between the intrinsic current
fluctuations.
\par
Now we have all requirements to calculate the shot noise in a
ferromagnetic multi-terminal diffusive structure. The first step
is to solve the diffusion equations (\ref{lapfi}-\ref{j})
implemented by the boundary conditions which are temporal current
conservation laws in the contacts to obtain the mean distribution
matrix in N metal and the charge and spin currents fluctuations in
the end points of N metal in terms of the Langevin current
fluctuations and the intrinsic current fluctuations at these
points. Now, knowing the correlations between these fluctuations
from Eqs. (\ref{jcjc}-\ref{jcis}) and (\ref{intcorrel}), enable us
to calculate the correlations of the charge and spin currents
fluctuations. In the next section we demonstrate the above
developed formalism by studying the noncollinear shot noise in a
two terminal spin-valve.


\section{Angular shot noise of a double barrier F-N-F system}

\begin{figure}
\centerline{\includegraphics[width=7cm,angle=0]{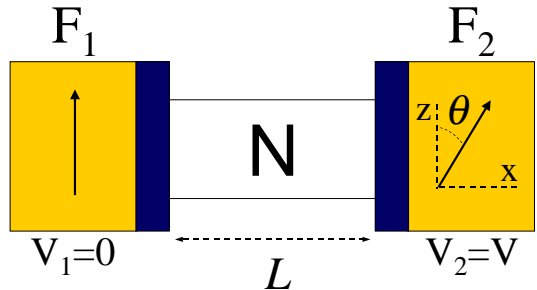}}
\caption{A schematic view of the system. A diffusive normal metal
wire connected to two ferromagnetic reservoirs through the tunnel
junctions. Ferromagnetic reservoirs are in thermal equilibrium and
magnetizations of them can orient in arbitrary directions.}
\label{system}
\end{figure}

We consider a two terminal diffusive spin-valve system as shown in
Fig.\ref{system}. Two ferromagnetic reservoirs ${\text F_1}$ and
${\text F_2}$ with a noncollinear orientation $\theta$ of the
magnetization vectors are connected through tunnel contacts to a
diffusive normal metal wire (N) of length $L$. The reservoirs
${\text F_{1,2}}$ are held at $0$ and $V$ potentials,
respectively. We consider a symmetric structure for which both of
the tunnel contacts have the same conductance and shot noise
matrices given by Eqs.(\ref{conductance}-\ref{noise}).
\par
The mean distribution function matrix inside the N metal is
obtained by solving the diffusion equation (\ref{lapfi}). The
general solution of this equation can be written as the energy
integral of the following relation:
\begin{eqnarray}&&
\bar{\hat{f}}_{\text{N}}(x)=\left[\hat{f_1}+(\hat{f_2}-\hat{f_1})
(a+b\frac{x}{L})\right]\hat{1} \nonumber
\\*
&&+\hat{\mbox{\boldmath $\sigma$}}\cdot\left[\vec{c}\sinh(\lambda
x/L)+\vec{d}\cosh(\lambda x/L)\right](\hat{f_2}-\hat{f_1}),
\label{avpotential}\hspace{0.5cm}
\end{eqnarray}
where the matrix distribution function at the terminals F$_{1,2}$
are determined by the Fermi distribution functions via
$\hat{f_1}=f_{\text{FD}}(\varepsilon)\hat{1}$ and
$\hat{f_2}=f_{\text{FD}}(\varepsilon-eV)\hat{1}$, and
$\lambda=L/\ell_{\text {sf}}$ is a dimensionless parameter which
measures the intensity of the spin-flip scattering. The scalars
$a$, $b$ and components of the vectors $\vec{c}=(c_1,c_2,c_3)$ and
$\vec{d}=(d_1,d_2,d_3)$ are coefficients which have to be
determined by the boundary conditions. The boundary conditions are
the spin-charge current conservation rules which in the absence of
spin-flip process at the tunnel contact $i$ ($i=1,2$), read
\begin{eqnarray}
\bar{\hat{I}}_i^{\text{C}}+\bar{\hat{I}}_i^{\text{N}}=0,
\label{dcconserv1}
\end{eqnarray}
Where $\bar{\hat{I}}_i^{\text{C}}$ and
$\bar{\hat{I}}_i^{\text{N}}$ are expressions for the mean matrix
current through the $i$th contact which are calculated by
relations (\ref{avcurrent}) and the diffusion equation (\ref{j}),
respectively. The mean charge current is determined by the
coefficient $b$ only:
\begin{eqnarray}
\bar{I}_{\text{c}}=Tr(\bar{\hat{I}})=-2G_{\text{N}}b,
\end{eqnarray}
where the normal conductance of the N wire is given by
$G_{{\text{N}}}=\sigma A/L$, $A$ is the cross section area of the
wire.
\par
To obtain correlations of the Langevin sources in Eqs.
(\ref{jcjc}-\ref{jcis}) we need the matrix distribution function
in the quantization axis parallel to the mean spin distribution
vector. This is achieved by diagonalizing the distribution
function matrix to get $\bar{f}_{\bar{s},\pm }(x)$. The result is
\begin{eqnarray}
\bar{f}_{\bar{s},\pm}(x)&=&f_1+(f_2-f_1)\left[(a+b\frac{x}{L})\right.
\nonumber
\\*
&&\pm\left.|\vec{c}\sinh(\lambda x/L)+\vec{d}\cosh(\lambda
x/L)|\right], \label{avpotential1}\hspace{0.5cm}
\end{eqnarray}
where $f_{1,2}$ are the corresponding Fermi distributions in the
terminals. Now we calculate the fluctuations of the matrix current
$\Delta \hat{I}^{\text{N}}$ at the contacts $(x=0,L)$ inside N
wire. Using the equation for $\delta \hat{\mathbf{J}}$ (\ref{j})
and Eq. (\ref{divj}) we obtain
\begin{eqnarray}
\Delta \hat{I}^{\text{N}}(0,L)=-\sigma\oint d\mathbf{s}\cdot
\left[ \nabla \phi_{{\text {c}}0(L)}(x)
\frac{Tr(\delta\hat{\varphi})}{2}\hat{1} \right. \nonumber
\\*
\left. +\nabla\phi_{{\text {s}}0(L)}(x)\left(\delta\hat{\varphi}-
\frac{Tr(\delta\hat{\varphi})}{2}\hat{1} \right)\right] +\delta
\hat{I}(0,L), \label{deli1}
\end{eqnarray}
where integration is over the surface of the N metal and
\begin{eqnarray}&&
\hspace{-0.8 cm}\delta \hat{I}(0,L)=\int
d\mathbf{r}\left[\frac{1}{2}\left(Tr(\hat{\mbox{\boldmath
$\eta$}})\cdot \nabla \right) \phi_{{\text {c}}0(L)}(x)\hat{1}
\right. \nonumber
\\*&&
\left.\hspace{-0.5 cm} +\left(\hat{\mbox{\boldmath $\sigma$}}
\cdot \mbox{\boldmath $\zeta$}+\left(\hat{\mbox{\boldmath
$\eta$}}- \frac{Tr(\hat{\mbox{\boldmath $\eta$}})}{2}\hat{1}
\right)\cdot\nabla\right)\phi_{{\text{s}}0(L)}(x)\right].
\label{delic}
\end{eqnarray}
Here the potential functions are defined as
\begin{eqnarray}&&
\phi_{{\text{c}}0}(x)=1-\frac{x}{L},
\\*&&
\phi_{{\text{c}}L}(x)=\frac{x}{L},
\\*&&
\phi_{{\text{s}}0}(x)=\frac{\sinh[\lambda(1-x/L)]}{\sinh(\lambda)},
\\*&&
\phi_{{\text{s}}L}(x)=\frac{\sinh(\lambda
x/L)}{\sinh(\lambda)}.\label{phcs}
\end{eqnarray}
Note that as a result of the spin-flip scattering the spin current
and hence, its fluctuations are not conserved through the wire.
\par
Now we can apply the current conservation law for the fluctuations
of the matrix currents at two contacts:
\begin{eqnarray}
\Delta\hat{I}_i^{\text{C}}+\Delta\hat{I}_i^{\text{N}}=0
\label{dcconserv}
\end{eqnarray}
Replacing the expressions of the current fluctuations from Eqs.
(\ref{deli1}-\ref{phcs}) and Eq. (\ref{dcc}) in Eq.
(\ref{dcconserv}) we obtain a system of eight linear equations
whose solutions give the fluctuations of the matrix chemical
potential $\delta\hat{\varphi}(0,L)$ at the points $x=0,L$ in
terms of the intrinsic matrix current fluctuations $\delta
\hat{I}_{1,2}$ and the Langevin matrix current fluctuations
$\delta \hat{I}(0,L)$. Replacing this result into Eq. (\ref{dcc})
the matrix current fluctuations are expressed in terms of the
fluctuating Langevin sources of currents and the intrinsic current
matrices as
\begin{eqnarray}
\Delta
\hat{I}^{\alpha\beta}&=&\sum_{\alpha^{\prime}\beta^{\prime}}
\left[A^{\alpha\beta}_{\alpha^{\prime}\beta^{\prime}}\delta\hat{I}^
{\alpha^{\prime}\beta^{\prime}}_1+B^{\alpha\beta}_{\alpha^{\prime}\beta^{\prime}}
\delta\hat{I}^{\alpha^{\prime}\beta^{\prime}}_2
\right.\nonumber\\*&&\left.
+C^{\alpha\beta}_{\alpha^{\prime}\beta^{\prime}}\delta\hat{I}^{\alpha^{\prime}\beta^{\prime}}(0)
+D^{\alpha\beta}_{\alpha^{\prime}\beta^{\prime}}\delta\hat{I}^{\alpha^{\prime}\beta^{\prime}}(L)\right],
\hspace{0.5cm} \label{deltai}
\end{eqnarray}
where $A^{\alpha\beta}_{\alpha^{\prime}\beta^{\prime}}$
,$B^{\alpha\beta}_{\alpha^{\prime}\beta^{\prime}}$,
$C^{\alpha\beta}_{\alpha^{\prime}\beta^{\prime}}$ and
$D^{\alpha\beta}_{\alpha^{\prime}\beta^{\prime}}$ are the
coefficients which depend on $G/G_{\text{N}}$, $P$, $\lambda$, and
$\theta$. From this result we can calculate the correlations of
the charge current $S=2\int dt\langle\Delta I_{\text{c}} \Delta
I_{\text{c}}\rangle$ and the corresponding Fano factor
$F=S/2e\bar{ I}_{\text{c}}$. The resulting expressions are too
lengthy to be written down here. In the next section we analyze
the magneto shot noise, the dependence of the $F$ on the the
relative angle $\theta$, for different values of the involved
parameters.


\section{results and discussion}

\begin{figure}
\centerline{\includegraphics[width=9cm,angle=0]{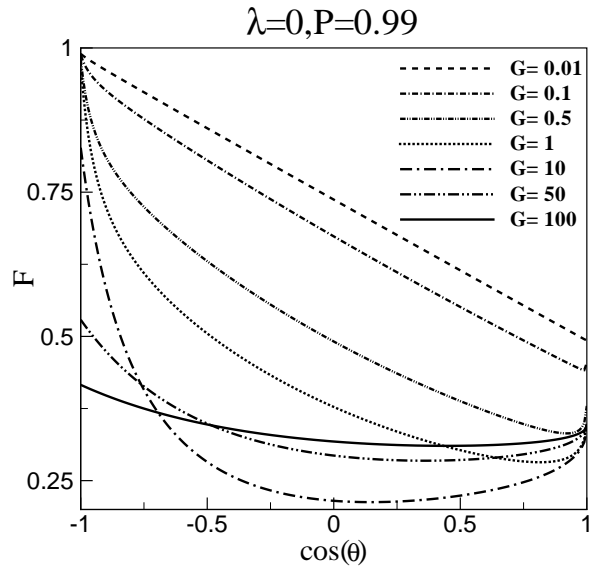}}
\caption{Fano factor versus the relative angle $\cos(\theta)$ for
different values of $G/G_{\text{N}}$ ($G_{\text{N}}=1$) in the
limit of $\lambda\rightarrow 0$ and for $P=0.99$.}
\label{fig2}
\end{figure}

Let us start our analysis of the shot noise with the limiting case
of negligible spin-flip scattering, {\it i.e.} when
$\lambda\rightarrow 0$, and highly resistive tunnel contacts
$G/G_{\text{N}}\rightarrow 0$.  In this limit we retrieve the
results of Tserkovnyak and Brataas for charge current shot
noise\cite{Tserkovnyak01}, that is, a monotonic variation of $F$
as a function of the relative angle of the magnetization vectors
$\theta$:
\begin{eqnarray}
F=\frac{1}{2}\left[1+P^2\sin^{2}(\theta/2)\right]. \label{fctb}
\end{eqnarray}
For a finite $G/G_{\text{N}}$ the angular dependence of Fano
factor deviates from the above simple monotonic behavior. This is
illustrated in Fig. \ref{fig2}, where we have plotted $F(\theta)$
for different values of $G/G_{\text{N}}$ in the limit
$\lambda\rightarrow 0$ and the polarization $P=0.99$. By
increasing $G/G_{\text{N}}$, $F$ decreases below the value given
by Eq. (\ref{fctb}) and finds a nonmonotonic dependence on
$\theta$. The nonmonotonic behavior is more pronounced for a value
of $G/G_{\text{N}}$ which depends on the polarization. The Fano
factor develops a minimum at a finite angle $\theta\neq 0$, which
depends on $P$ and $G/G_{\text{N}}$. In the limit of
$G/G_{\text{N}}\gg 1$, $F$ returns back to the constant normal
state value $1/3$, as it is expected. This holds for not perfectly
polarized terminals $P\neq1$. We found that for $P=1$, Fano factor
retains its nonmonotonic angular dependence in the limit
$G/G_{\text{N}}\gg 1$. In this case $F$ has a minimum well below
$1/3$ at a finite $\theta$, a sharp peak at $\theta=\pi$ where it
reaches one and a smooth peak at $\theta= 0$ with the value $1/3$.
The corresponding behavior in the conductance of the system was
studied in Ref. \onlinecite{Huertas00}.
\par
\begin{figure}
\centerline{\includegraphics[width=9cm,angle=0]{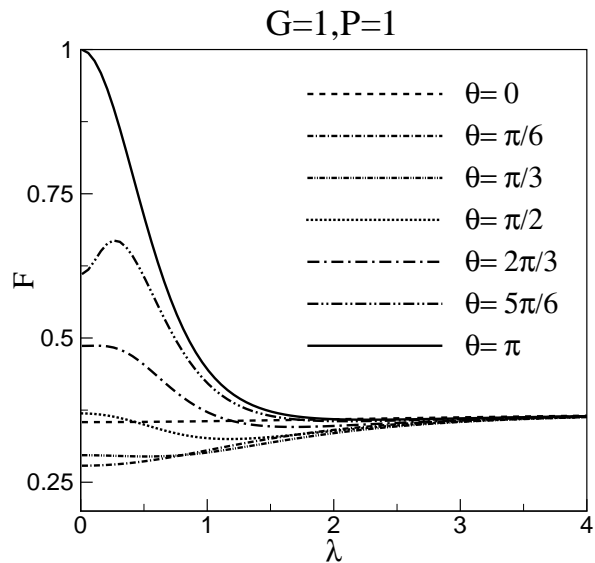}}
\caption{Fano factor versus the spin-flip scattering intensity
$\lambda=L/\ell_{sf}$ for different values of relative angle
$\theta$ for a system with $G=1$ and $P=1$.}
\label{fig3}
\end{figure}

Next we study the effect of spin-flip scattering in the N metal.
In Fig. \ref{fig3} we show dependence of Fano factor on $\lambda$
for different angles, when $G=1$ and $P=1$. As it can be seen from
Fig. \ref{fig3}, a strong spin-flip scattering destroys any spin
polarization and thus suppresses the dependence of the Fano factor
on $\theta$. In this limit ($\lambda\gg 1$) $F$ reduces to the
normal state value independent of $\theta$
\begin{eqnarray}
F=\frac{1}{3}\frac{12+12G/G_{\text{N}}+6(G/G_{\text{N}})^2+
(G/G_{\text{N}})^3}{(2+G/G_{\text{N}})^3}. \hspace{0.3
cm}\label{allnormalv}
\end{eqnarray}
At a finite $\lambda$ the shot noise of different $\theta$s are
separated. The most strong variations happens for
$\lambda\rightarrow 0$. Increasing $\lambda$ from zero,
$F(\lambda)$ passes through a minimum or maximum, depending on
$\theta$, before reaching the normal state value at large
$\lambda$s. In the case of Fig. \ref{fig3} the minimum and maximum
happen for $\theta<\pi/2$ and  $\theta>\pi/2$, respectively. While
the minimum and maximum values of $F$ are located at $\lambda=0$
for $\theta=0$ and  $\theta=\pi$, they can happen at finite
$\lambda$s for intermediate angles.
\par
\begin{figure}
\centerline{\includegraphics[width=9cm,angle=0]{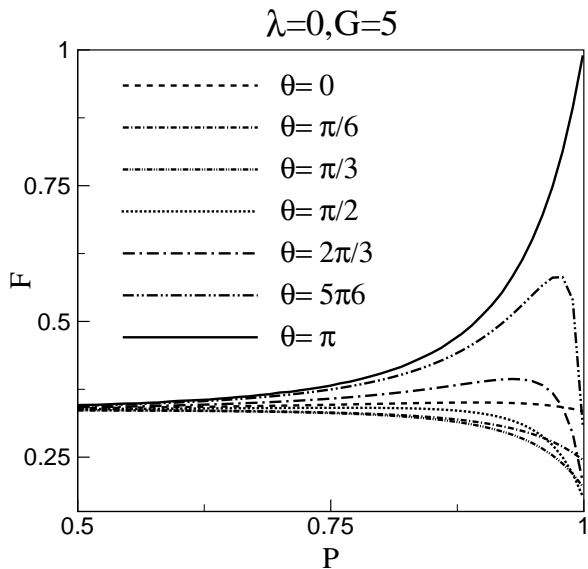}}
\caption{Fano factor versus polarization of contacts $P$ for
different values of $\theta$ in the limit of small spin-flip
scattering intensity $\lambda=L/\ell_{sf}\rightarrow 0$.}
\label{fig4}
\end{figure}

To show the effect of contacts polarization $P$, we have plotted
in Fig. \ref{fig4} the Fano factor vs. $P$ in the limit $\lambda
\rightarrow 0$, for different $\theta$s. For $P\rightarrow 0$ the
Fano factor takes the normal state value (\ref{allnormalv}),
irrespective of the value of $\theta$. Increasing $P$ the
dependence of $F$ on $\theta$ is appeared. The Fano factor
increases (decreases) monotonically with $P$ for $\theta=\pi$
($\theta=0$) and reaches a maximum (minimum) at $P=1$. However for
an intermediate angles $F(P)$ can have a nonmonotonic variation
with a maximum at $P< 1$. This is in contrast to the limit of the
normal metal node geometry ($G/G_{\text{N}}\rightarrow 0$), where
a monotonic dependence on $P$ was predicted (Eq. (\ref{fctb})).
%


\section{conclusions}

In this paper we have presented a semiclassical theory of the
spin-polarized current fluctuations in a diffusive normal metal
which is connected by tunnel contacts to ferromagnetic terminals
with noncollinear magnetizations. Based on the Boltzmann-Langevin
approach, we have developed diffusion equations which allow for
calculation of the charge-spin distribution and current density
matrices and the correlations of the corresponding fluctuations in
noncollinear multiterminal systems. Our theory takes into account
the precession of the noncollinear spin accumulation as well as
the spin-flip induced relaxation in the normal metal. Applying the
developed theory to a two terminal FNF structure we have found
that the Fano factor has a nonmonotonic dependence on the
magnetizations angle, provided that the spin-flip intensity is
small and the tunnel contacts have an appreciable polarization.
For antiparallel orientation the Fano factor is found to increases
well above the unpolarized value due to a large spin accumulation
in the N metal. In contrast, for the parallel orientation the shot
noise is almost identical with its normal state (unpolarized)
value because of the spin accumulation suppression. At the
intermediate angles we have found that the interplay between the
spin accumulation precession and the suppression with
magnetizations angle causes that the Fano factor develops a
minimum. The minimum shot noise can be substantially below the
normal value depending on the relative conductances of the N metal
and tunnel contacts $G/G_{\text{N}}$ and the polarization $P$. We
further have shown that the spin-flip scattering diminishes the
amplitude of the nonmonotonic behavior as well as the polarization
effects.
\par
In spite of a few early \cite{Moodera96,Nowak99} and recent
\cite{Tserkovnyak06} experimental studies devoted to
spin-polarized shot noise in magnetic tunnel junctions between
ferrromagnets, to our knowledge, there have not been reports on
shot noise measurements in metallic spin-valve systems. We expect
that such experiments will be performed in near future, which will
reveal the effects predicted in the present paper.\\


\begin{acknowledgments}
We acknowledge useful discussions with G. E. W. Bauer and W.
Belzig.
\end{acknowledgments}

\end{document}